\documentclass[conference]{IEEEtran}

\usepackage[pdftex]{graphicx}
\graphicspath{{../pdf/}{../jpeg/}}
\DeclareGraphicsExtensions{.pdf,.jpeg,.png}

\usepackage[cmex10]{amsmath}
\usepackage{mathabx}
\usepackage{algorithmic}
\usepackage{array}
\usepackage{mdwmath}
\usepackage{mdwtab}
\usepackage{eqparbox}
\usepackage{url}
\usepackage[draft]{hyperref}

\begin{document}

\title{Drowsiness Detection for Office-based Workload with Mouse and Keyboard Data}


\author{
    \IEEEauthorblockN{
        Sanurak~Natnithikarat\IEEEauthorrefmark{1},
        Sirakorn~Lamyai\IEEEauthorrefmark{2},
        Pitshaporn~Leelaarporn\IEEEauthorrefmark{3},
        Narin~Kunaseth\IEEEauthorrefmark{3},
        Phairot~Autthasan\IEEEauthorrefmark{3},\\
        Thayakorn~Wisutthisen\IEEEauthorrefmark{4}
        and Theerawit~Wilaiprasitporn\IEEEauthorrefmark{3}
    }
    \IEEEauthorblockA{
        \IEEEauthorrefmark{1}Ruamrudee International School
    }
    \IEEEauthorblockA{
        \IEEEauthorrefmark{2}Department of Computer Engineering, Kasetsart University
    }
    \IEEEauthorblockA{
        \IEEEauthorrefmark{3}School of Information Science and Technology, Vidyasirimedhi Institute of Science and Technology\\theerawit.w [at] vistec.ac.th
    }
    \IEEEauthorblockA{
        \IEEEauthorrefmark{4}School of Information Technology, King Mongkut’s University of Technology Thonburi
    }
}

\maketitle

\begin{abstract}
Non-invasive devices involved in the detection of drowsiness generally include infrared camera and Electroencephalography (EEG), of which sometimes are constrained in an actual real-life scenario deployments and implementations such as in the working office environment. This study proposes a combination using the biometric features of keyboard and mouse movements and eye tracking during an office-based tasks to detect and evaluate drowsiness according to the self-report Karolinska sleepiness scale (KSS) questionnaire. Using machine learning models, the results demonstrate a correlation between the predicted KSS from the biometrics and the actual KSS from the user input, indicating the feasibility of evaluating the office workers' drowsiness level of the proposed approach.
\end{abstract}


\section{Introduction}
\IEEEPARstart{D}{rowsiness} has been one of the persistent issues in an office-like environment in all areas of industry, leading to occupational fatigue from overworking, especially in a scenario where high accuracy of operation output is required. Modern companies are now aiming with ample efforts to monitor its workplaces in order to maintain the employees’ optimal working condition and ensure a satisfying level of productivity. 

With the growing advanced technology, various physiological techniques have been adopted for drowsiness detection, such as the measurement of eye movements requiring the additional installations of infrared (IR) cameras, electroencephalography (EEG), electrooculography (EOG), and hybrid brain--computer interfaces (BCIs) \cite{lal2001critical, Kurylyak2011, Wilaiprasitporn2016}. However, some of these devices entail the attachment of electrodes on the body parts such as the scalp and the skin \cite{Pal2008, Sone2013}, causing discomforts to the employees which may be considered as infeasible for the real-life applications. For this reason, different methods have been established without interfering with the working routines. A system to monitor mental fatigue and behavior featuring the usage of computer set including the keyboard and the mouse has been studied with the goal to improve the working environment \cite{pimenta2013monitoring}. The fine control and objects interaction patterns have been demonstrated to be altered significantly when different levels of drowsiness were reported. Furthermore, a convenient, though subjective report, questionnaire known as Karolinska sleepiness scale (KSS) has been validated against the brain signal activity detected by EEG and EOG and has been widely implemented to assess drowsiness \cite{Kaida2006}. The KSS consists of a 9-point scale, ranging from 1 for very alert to 9 for very sleepy, with 5 as neither alert nor sleepy.

In the present study, we examine the changes in the keyboard and mouse usage patterns to create a predictive model in relation to the Karolinska Sleepiness Scale (KSS) \cite{Kaida2006} to approximate and classify the level of drowsiness of the office workers. We also seek for the possibility of combining the proposed computer approach with a traditional camera-based method to create a drowsiness monitoring system in a comfortable office-like environment.

\section{Experiment}
\subsection{Participants}
The total of 18 healthy participants (age 21-35; 15 males and 3 females) in this study were recruited from a pool of students and faculty members at Vidyasirimedhi Institute of Science and Technology, Thailand. All participants were instructed to finish a meal prior to the beginning of the trials and advised against the increase of the caffeine intake more than the usual amount each of the subject consumes in a normal weekday. The trials were conducted in the afternoon after the lunch break. Each participant was given an adequate period of time to rest after their meal. The informed consents were obtained from all participants following the Helsinki Declaration of 1975 (as revised in 2000).

\begin{figure*}[tbp]
    \includegraphics[width=\textwidth]{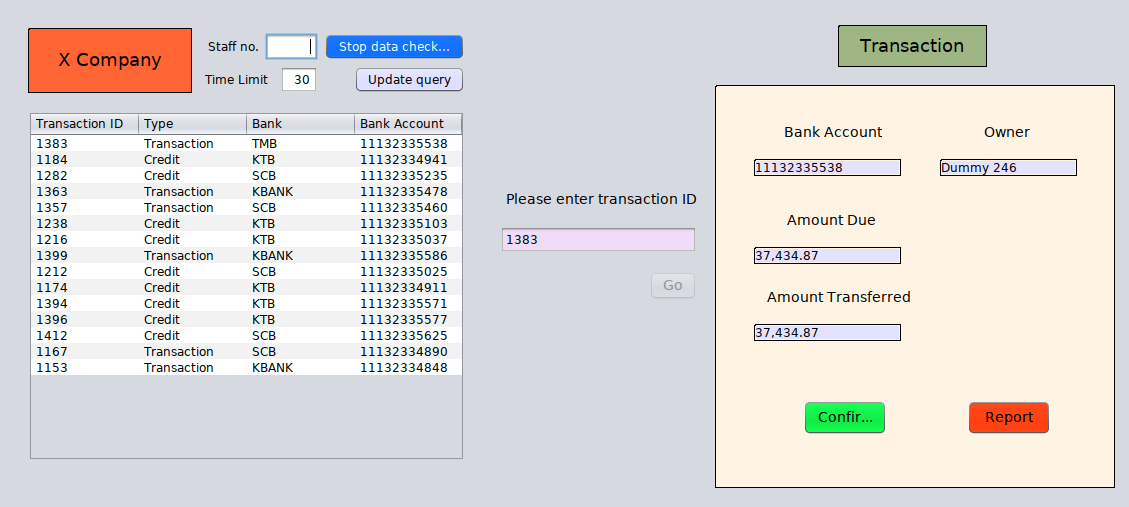}
    \caption{A screenshot of the application with the designed task load. In this state, the participants will need to verify the transaction correctness and decide to confirm or report the transaction.}
    \label{drowsiness-app-tx-noborder}
\end{figure*}

\subsection{Experimental Design}

A set of tasks was designed to simulate a working routine in a company's accounting division. The designed tasks were adapted from a scenario of an actual task load of the accountants, monitoring the financial transactions. The interface of the program simulating such task load consists of two main parts: a list of transactions and a verification area. The list contains the information including of the transaction ID, payment method, the name of the bank, and the bank account number of each customer. The verification area comprises of dialog boxes for the name of the payer, bank account number, the amount due and the transferred amount. The participants were instructed to select one transaction, enter the ID, and click `Go' button to proceed with the payment verification process. The details in the verification area will subsequently appear. The main task was to verify if (1) the amount due and the amount transferred matched and (2) the bank account in the transaction list and in the verification area matched. The participants were instructed to choose `confirm' if all information were correct or `report' if a part of the information is false. To complete the verification task, a username and password prompt appeared, of which the participants were asked to enter the correct set given prior to the beginning of the experiment. At the end of each transaction, the participants were inquired to rate their confidence from 1 to 10 for each transaction verified. The experiment lasted for approximately one hour. Furthermore, a pop-up KKS questionnaire with labels indicated in \autoref{kss-phrases} would appear randomly every three to eight minutes during the experimental period and the participants were inquired to rate their level of drowsiness at the exact moment.

\begin{table}
    \centering
    \caption{Description of the Karolinska Sleepiness Scale}
    \label{kss-phrases}
    \begin{tabular}{c|l}
        \hline
        \textbf{Level}  & \multicolumn{1}{c}{\textbf{Description}}\\
        \hline
        1               & Extremely alert \\
        3               & Alert, normal level \\
        5               & Neither alert nor sleepy \\
        7               & Sleepy, but no effort to keep alert\\
        9               & Very sleepy, great effort to keep alert, fighting sleep\\
        \hline
    \end{tabular}
\end{table}

\subsection{Participant stimulation}

The experiment was conducted in a well-lit and isolated room to control the effects of lights, sounds, and other forms of disturbance that might occur. A set of computer, consisting of a 27-inch LED monitor, a keyboard, and a mouse were installed in the room. Each of the participants was allowed to adjust the chair and reposition the monitor's height to the suitable working postures. The overall experimental process lasted for one hour.

\subsection{Data Acquisition}

The behavioral features for the detection and classification of drowsiness were logged from the mouse movements and keyboard strokes during the experiment as well as the decisions to confirm or report the transaction along with the actual correct answer to evaluate the performance of the participants. In this study, the considered biometric features include key down time, the number of delete button pressed, the mouse average error, and the mouse velocity.

\subsection{Data Analysis}

The distribution of the actual KKS score over the predicted KSS score were performed using linear regression and were converted by using the principal component analysis (PCA) into acceptable variables. Support Vector Machine (SVM) was then used to predict the estimated KSS from the biometric features from the participants’ behavior.

\section{Results}
\subsubsection{Analysis of KSS}
\begin{figure}
    \centering
    \includegraphics[width=1\columnwidth]{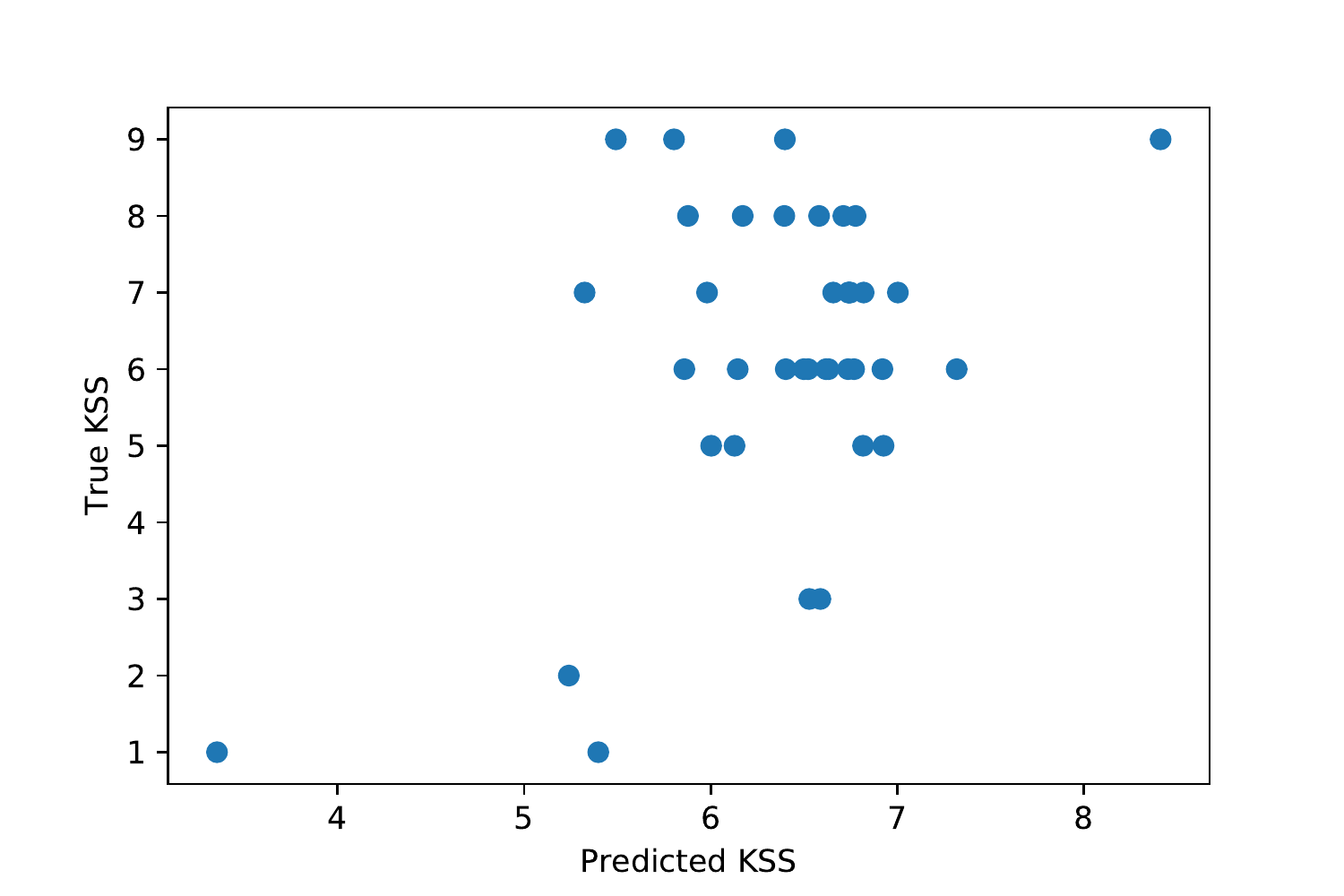}
    \caption{A linear regression displaying the regular KSS scores.}
    \label{Regular_KSS}
\end{figure}

The results of simple linear regression analysis as shown in \autoref{Regular_KSS} demonstrated that the distribution of the KSS score is skewed left. Setting each of the participant’s initial as the anchor KSS score, each score was subtracted by the initial value. This difference of KSS score was evaluated and plotted on against the frequency. The frequency plot is shown to be similar to a normal distribution with a mean of 0.95 and standard deviation of 2.63. 

\begin{figure}
    \centering
    \includegraphics[width=1\columnwidth]{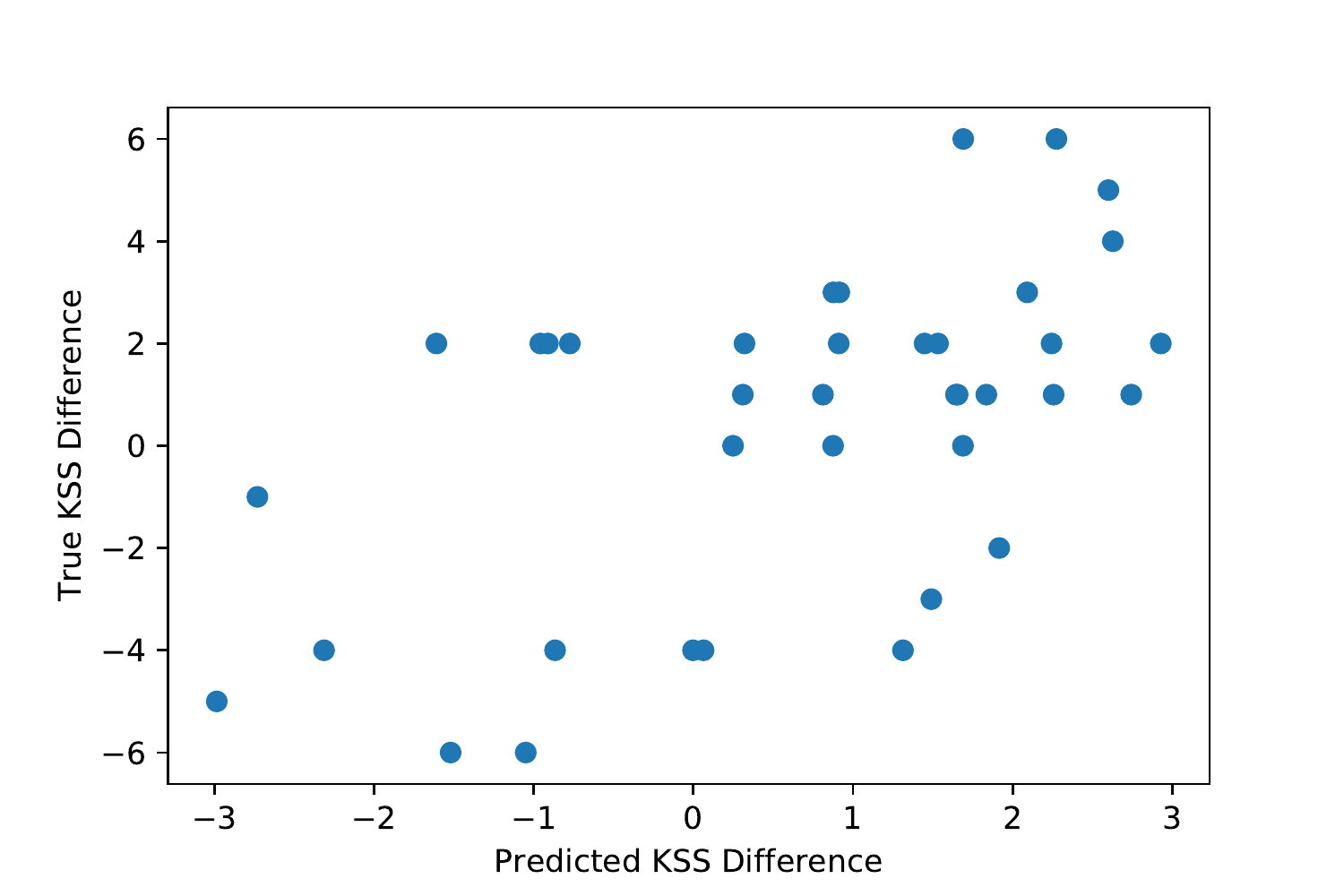}
    \caption{Differences in the distribution of the true KSS and the predicted KSS score.}
    \label{Diff_KSS}
\end{figure}


\autoref{Diff_KSS} displays a general upward trend in the plot of the KSS difference against time. The correlation coefficient was found to be 0.69. As the experiment continued, the difference of KSS was revealed to be diving into more positive value.



\subsubsection{Linear Regression}
The linear regression model used for features included the mouse average error, the mouse velocity, the frequency of pressing the delete button, and key down time. From these features, the linear regression model gave the following parameters: 0.00100983, 0.0006003, -0.08821249, and -0.03016235, respectively. Some features were randomly removed in order to test the significance of each variable. Removing the mouse average error, the mouse velocity, the frequency of pressing the delete button, and key down time, the correlation coefficients yielded are 0.55, 0.24, 0.47, and 0.50, respectively. The features were plotted against the KSS difference. The correlation coefficients were found to be 0.24, 0.078, -0.5, and -0.009, in the same order. The calculated correlation was -0.5 and has been shown to be significant compared to the other features. 

\subsubsection{Principal Components Analysis}
Dimension reduction through the PCA was conducted in order to reduce the complexity of the linear regression analysis. Using PCA, the dimension of the data was reduced from 4 to 3. Retraining the linear regression model on the three reduced features and repeating the same testing procedure, the R-value was found to be 0.55. After replicating this with the regular KSS score, the correlation coefficient dropped to 0.31.

\begin{figure}
    \centering
    \includegraphics[width=1\columnwidth]{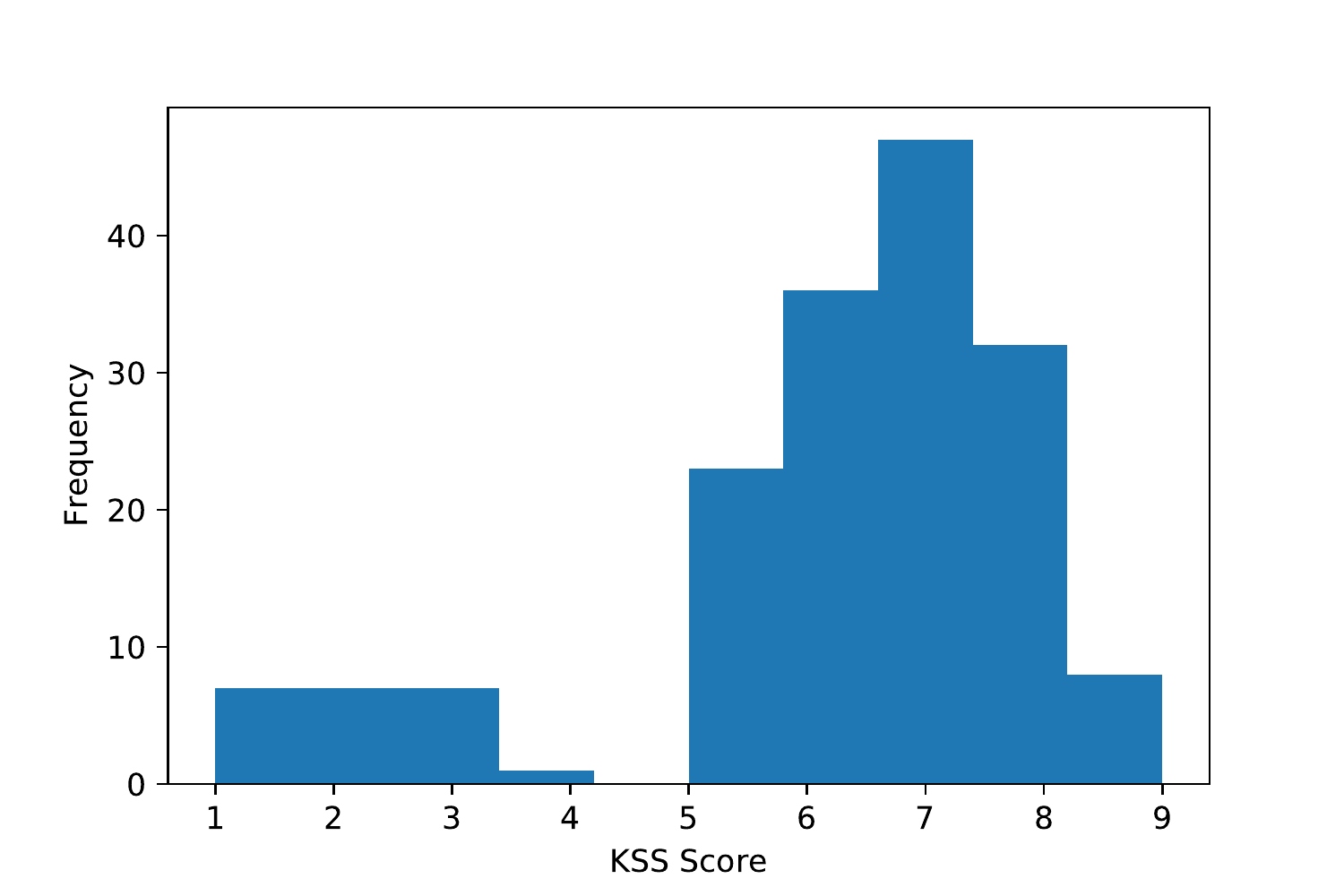}
    \caption{The distribution of the true KSS and the predicted KSS score. Fitting this distribution into normal distribution results in $mu = 0.95$, $SD=2.63$}
    \label{KSS_Dis}
\end{figure}

\subsubsection{Support Vector Machine}
Using the same testing methods as the linear regression analysis, a SVM model was trained. The algorithm yielded an R-Value of 0.70. The performance of the SVM produced higher quality than the linear regression because the SVM algorithm is not affected by the outliers in the data, whereas the linear regression model was. When the regular KSS score was used as the label, the SVM performance dropped to 0.36.

\begin{figure}
    \centering
    \includegraphics[width=1\columnwidth]{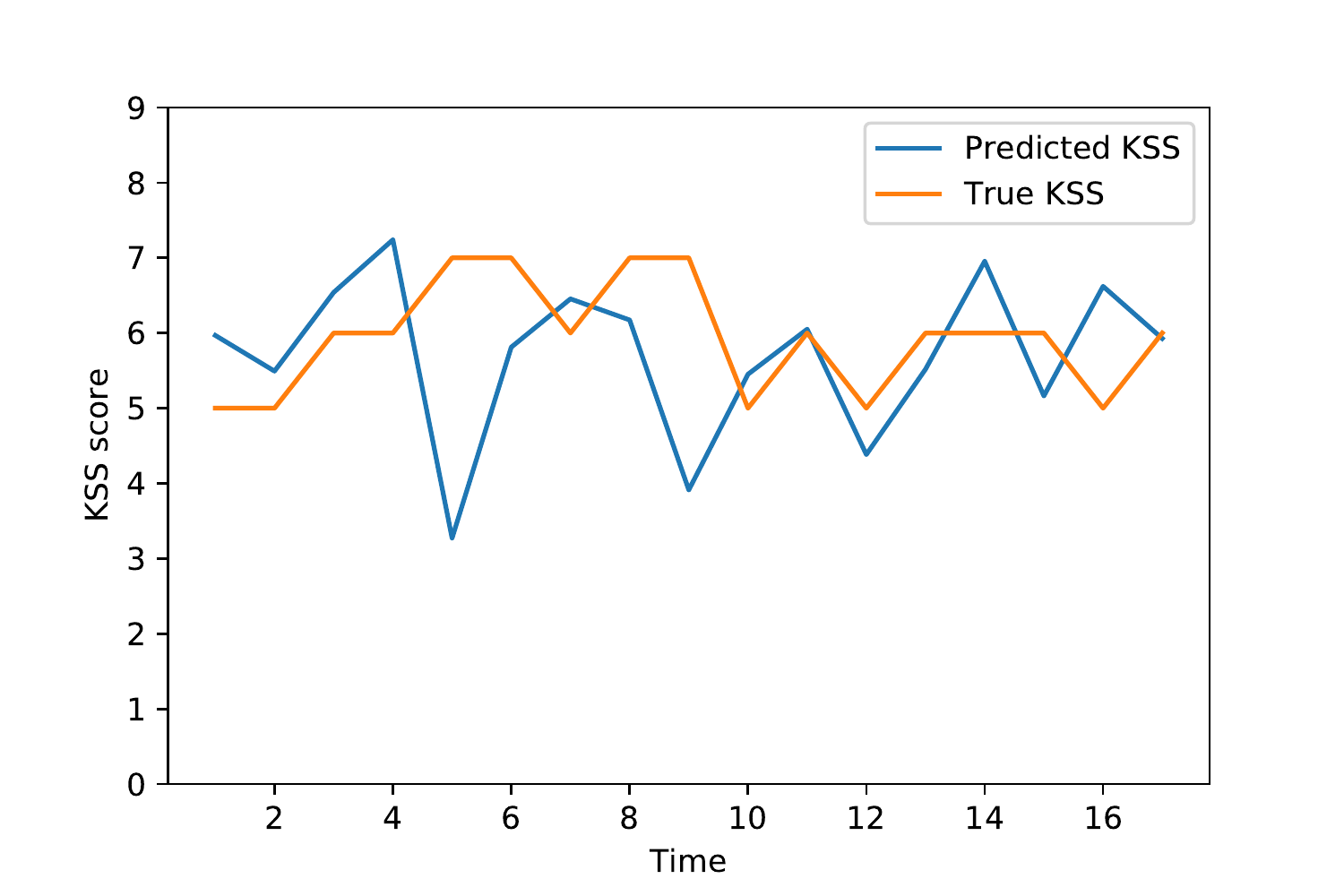}
    \caption{Comparison of the predicted KSS difference and actual KSS difference through time of a single subject using the trained SVM model on a testing set.}
    \label{predicted-kss-through-time}
\end{figure}

The prediction of KSS through time of a single subject, comparing to the actual test data is as shown in \autoref{predicted-kss-through-time}. It could be observed that the predicted KSS using only mouse and keyboard data follows well along the actual KSS rated by the subjects.

\begin{figure}
    \centering
    \includegraphics[width=1\columnwidth]{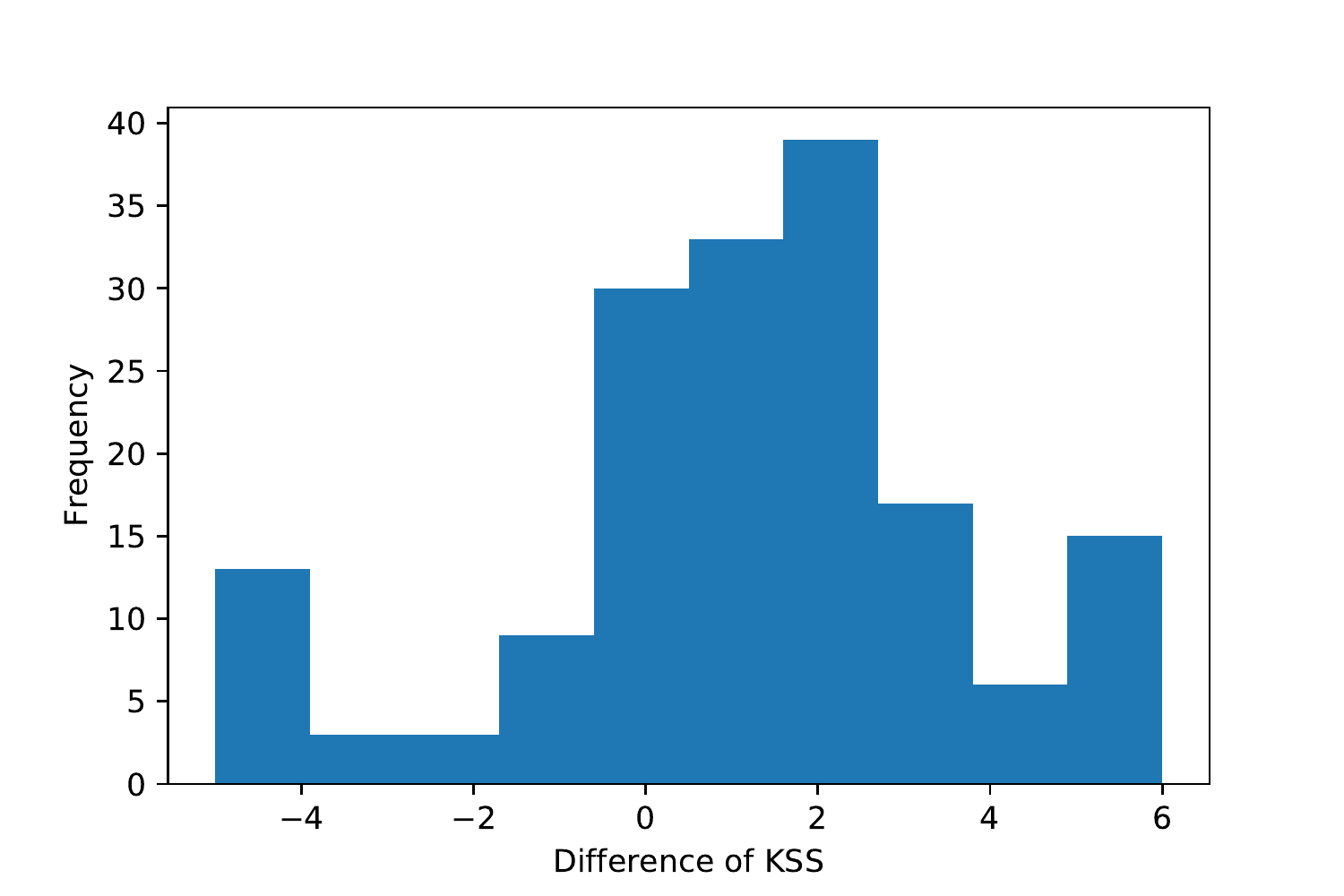}
    \caption{The distribution of the difference between the true KSS and the predicted KSS score.}
    \label{KSS_Diff_Dis}
\end{figure}

\section{Discussion}
\subsubsection{KSS Difference}
Although the definition of drowsiness and fatigue can be subjective and viewed differently amongst the participants, it is required that we utilize the difference of the KSS score as our data in order to detect the changes rather than pinpointing the participant's self-rated sleepiness score. As mentioned in the Result Section, the distribution of the KSS score is displayed as skewed left, indicating that most of the participants rated themselves according to their perception of being drowsy during the experiment. When using the KSS score instead of the KSS difference, every model's effectiveness drops significantly. Using the initial KSS score of the participant as a basis to which we compared the following KSS score reports, the participant-to-participant difference is minimized, yielding a more effective model. 

Choosing the two features with the highest correlation with the KSS score difference, we created another linear regression model. The new regression model received a correlation coefficient of 0.53 between the predicted KSS difference and the real KSS difference. In the linear regression, the keyboard down time did not improve the accuracy of the model. However, the frequency of pressing the delete button appears to be an important feature.

\subsubsection{SVM}
Using the same basis, the key down time feature was removed. The SVM algorithm correlation coefficient between the predicted and the true KSS coefficient dropped from 0.70 to 0.58. This indicates that the SVM algorithm is capable of finding a meaningful usage of the key down time feature, in which the linear regression model was not capable of. 

\subsubsection{Future Works}
Using the proposed experimental task, we hope to study fatigue and new biometrics for detecting drowsiness in an office--like environment. The usage of the experimental task replaces the need for expensive non-portable hardware and software when studying fatigue in drivers. With this in mind, we could evince the future studies to improve the approach and evince it to the real--world scenarios, especially drowsiness during driving. 

\section{Conclusion}
The improvements in the approach of a drowsiness detection system using the mouse movement and keyboard keystroke data as well as the self-evaluation of KSS could serve as an imperative opportunity to evaluate the drowsiness in an office-like environment. In this present study, we trained our data in order to predict the KSS score according to the collected bio—metric behaviors. Here, we have demonstrated that there is a strong correlation between the bio—metric features and the difference of KSS through time. This possibility of detecting the office workers' drowsiness level using the proposed approach could be developed to be applied for monitoring mental fatigue in a more realistic scenario in the near future. 

\appendices

\section{Drowsiness Task Load Application}

The application simulating the task load used in this paper can be obtained from \url{https://github.com/IoBT-VISTEC/drowsiness-taskload}.
\bibliographystyle{IEEEtran}
\bibliography{bibtex/bib/references.bib}

\end{document}